\documentclass[intlimits,twoside,a4paper]{article}

\usepackage{amsmath,amssymb}
\usepackage{graphicx}
\usepackage{epsfig}
\usepackage{color}

\usepackage[T2A]{fontenc}
\usepackage[cp1251]{inputenc}

\usepackage{cmpj2}

\issue{2015}{18}{2}{23702}
\doinumber{10.5488/CMP.18.23702}

\DeclareMathOperator{\Ai}{Ai}

\def\qqq{{\bf q}}
\def\kkk{{\bf k}}
\def\kkkp{{{\bf k}^\prime}}
\def\rrr{{\bf r}}

\title{Electron-acoustic phonon field induced tunnel scattering}

\author{S.V. Melkonyan, A.L. Harutyunyan, T.A. Zalinyan}

\address{Department of Physics of Semiconductors \& Microelectronics, Yerevan State University, 1 Alex Manoogian St.,
0025 Yerevan, Armenia}

\date{Received October 16, 2014}

\begin{document}

\maketitle

\begin{abstract}
Theory of electron-acoustic single phonon scattering has been reconsidered. It is assumed that the non-dege\-ne\-rate semiconductor has a spherical parabolic band structure. In the basis of the reconsideration there is a phenomenon of the tilting of semiconductor bands by the perturbing potential of an electric field. In this case, electron eigenfunctions are not plane waves or Bloch functions. In low-field regime, the expressions for electron intraband transition probability and scattering time are obtained under elastic collision approximation. Dependencies of scattering time on electron energy and uniform electric field are analyzed. The results of corresponding numerical computations for n-Si at $300$~K are presented. It is established that there is no fracture on the curve of electron scattering time dependence on the electron energy.
\pacs 72.10-d, 63.20.kd
\end{abstract}

%%%%%%%%%%%%

\section{Introduction}

Current carrier (electron) mobility $\mu$ is an important parameter characterizing many transport phenomena in semiconductors under electric field $F$. Electron mobility is determined as $\mu=\langle e\tau_\kkk /m\rangle$ \cite{1, 2}, where $\tau_\kkk$ is the electron quasi-momentum relaxation time, $e$ is the electron charge magnitude, $m$ is the electron effective mass, $\langle\cdots\rangle$ is the symbol of averaging over conduction zone quantum states. Relaxation time $\tau_\kkk$ is determined by the electron scattering by various dynamic and static imperfections of a crystall lattice such as lattice vibrations (optic and acoustic phonons), ionized and neutral impurity atoms, vacancies,  etc. For theoretical consideration of scattering probability and relaxation time $\tau_\kkk$, a flat-band semiconductor model is used, as a rule \cite{1, 2, 3, 4}. In this case, in low electric field region ($F<F_{\text{c}}$, where $F_{\text{c}}$ is a characteristic field), the electron relaxation time $\tau_\kkk$  and, therefore, the mobility $\mu$ are field-independent quantities \cite{1, 4, 5}. Particularly, the relaxation time related to electron-acoustic single phonon elastic scattering in non-degenerate n-type semiconductor with a spherical parabolic conduction band is given by \cite{1,2,3,4}
\begin{equation}\label{eq-1}
1/ \tau_{\kkk, \text{ac}}=\frac{D^2_{\text{ac}}(2m)^{3/2}k_{\text B} T}{2\pi \hbar^4 \rho_{\text r} \nu^2_0} \sqrt{\varepsilon_\kkk}\,.
\end{equation}
Here, $\rho_{\text r}$ is a reduced mass density of a crystal, $D_{\text{ac}}$ is the acoustic
deformation potential constant, $k_{\text{B}}$ is the Boltzmann constant, $T$ is temperature,
$\nu_0$ is the long-wavelength longitudinal acoustic phonon velocity,
$\varepsilon_\kkk=\hbar^2 k^2 /2m$ is the electron energy, boldface $\kkk$ is the electron wave vector
(herein below, the magnitudes of vector quantities are denoted by non-boldface symbols).

At high electric fields ($F>F_{\text{c}}$), the time $\tau_\kkk$  and, therefore, the mobility $\mu$
depend on the applied electric field \cite{4}. Thus, at electron scattering by an acoustic phonon,
electron mobility decreases with an increase of electric field above $F_{\text{c}}$ \cite{5}.
The magnitude of characteristic field $F_{\text{c}}$ depends on semiconductor parameters such
as crystallographic directions, impurity concentration, temperature, etc. It is of an order of
$10^3$~V/cm at $300$~K, e.g., for pure Si $F_{\text{c}} \approx 10^3$~V/cm, for pure Ge $F_{\text{c}}\approx 600$~V/cm,
for high purity GaAs $F_{\text{c}}\approx 2.8\cdot 10^3$~V/cm \cite{5}. The dependence $\mu(F)$ is explained
by the phenomenon of electron gas heating-up under the effect of a high electric filed \cite{1, 4}. However,
in recent work \cite{6}, a new mechanism of electron lattice scattering, referred to as electron-phonon FIT
(field induced tunnel) scattering, is observed. In the basis of the FIT scattering there lies a phenomenon
of tilting of semiconductor bands by the perturbing potential of an electric field. The effect of the
electron-phonon FIT scattering is explained in terms of penetration of an electron wave function into
a semiconductor band gap in the presence of an electric field. Contrary to a flat-band semiconductor,
in a tilted-band semiconductor, a conduction electron transition in the band gap region is allowed.
In   \cite{6}, reconsidering the electron-phonon interaction theory, the case of electron intraband FIT
scattering by non-polar optical phonon is analyzed. In the present work, electron-acoustic phonon FIT
intraband scattering is considered. It is assumed that the non-degenerate n-type semiconductor has a parabolic conduction band.

%%%%%%%%%%%%%%%%%%%%
\section{Electron-acoustic phonon FIT transition probability}

 To theoretically characterize the carrier scattering, it is necessary to consider the scattering probability and evaluate the relaxation time. The task of the transition probability calculation is solved based on the perturbation theory (see, e.g.~\cite{1,2,3,4}). According to this theory, the probability per unit time of quantum system transition from $\lambda$ state to $\lambda^\prime$ state, to the first order in the perturbation, is determined as \cite{1,2}
 \begin{equation}\label{eq-2}
W(\lambda, \lambda^\prime)=\frac{1}{\hbar^2}\frac{\text{d}}{\text{d}t}\left\vert \int\limits^t_0 a_{\lambda^\prime, \lambda}(t)\, \text{d}t\right\vert^2 .
\end{equation}
Here, $a_{\lambda^\prime, \lambda}(t)$ is the perturbation matrix element:
\begin{equation}\label{eq-3}
a_{\lambda^\prime, \lambda}(t)=\int\limits_V \text{d}{\bf R} \psi^*_{\lambda^\prime} ({\bf R}, t) \hat{W}({\bf R}, t) \psi_{\lambda} ({\bf R}, t),
\end{equation}
$\psi_{\lambda} ({\bf R}, t)= \psi_{\lambda} ({\bf R})\exp (-\text{i}E_{\lambda}t/\hbar)$, $\lambda$ is the set of quantum numbers characterizing different states of a non-perturbed system, $\psi_{\lambda} ({\bf R})$ and $E_{\lambda}$ are the wave function and energy eigenvalues of stationary state of a non-perturbed quantum system, respectively, $\hat{W}({\bf R}, t) $ is the perturbation operator, ${\bf R}$ is the set of the quantum system coordinates,  $V$ is the volume, `$*$' is the complex conjugate symbol.

Consideration of the electron scattering by phonons is based on \eqref{eq-2} and \eqref{eq-3} as well. In brief, the description of the probability calculation is as follows. In the present case, the quantum system consists of a conduction electron in the crystall periodic field and lattice normal vibrations. Then, $\lambda$ should
be replaced by an electron quasi-wave vector $\kkk$ and by the phonon occupation numbers of all possible states.  The electron-phonon interaction Hamiltonian $H_{\text{e-ph}}$ is taken as perturbation $\hat{W}({\bf R}, t)$;  the $\rrr$ radius vector and the normal coordinates of lattice vibration are taken as a quantum system coordinate ${\bf R}$. The wave function of non-perturbed state of an electron-phonon system is expressed as a product of one-electron wave function and harmonic oscillator wave functions.

In flat-band semiconductors, an electron state is described by Bloch functions.
Therefore, for an electron scattered from an initial state $\vert \kkk\rangle$ to
a final state $\vert \kkkp\rangle$, the transition probability per unit time $W(\kkk, \kkkp)$
is evaluated based on the  Bloch functions \cite{1,2,3,4}. To simplify the calculations,
a plane wave $\psi_\kkk(\rrr)=\re^{\text{i} \kkk\rrr}/ \sqrt{L_x L_y L_z}$ (where $L_x$, $L_y$
and $L_z$ are the sizes of a semiconductor, $V=L_x L_y L_z$) is used sometimes as an electron
wave function \cite{1,2,3,4} (nearly free electron approximation). Calculations of the
probability $W(\kkk, \kkkp)$  based on the Bloch function or plane wave are well known
and reported in detail in numerous publications (see, for example, \cite{1,2,3,4}).
However, in the present work, a semiconductor, whose bands are tilted by the perturbing potential of a uniform electric field, is of interest. In this case, electron eigenfunctions are not plane waves or Bloch functions \cite{1,3,7}. Here, based on such an assumption, the probability $W(\kkk, \kkkp)$  is recalculated following the above-mentioned general approach. It is assumed that the perturbation Hamiltonian $H_{\mathrm{e-ph}}$, which is a harmonic function of time (harmonic perturbation), is determined using the deformation potential theory \cite{1,2,3,4}  of electron-phonon interaction. The results of our calculations show that:
\begin{itemize}
\item
the probability of electron transition with phonon absorption can be presented as follows:
\begin{equation}\label{eq-4}
W_a(\kkk, \kkkp)= \frac{D^2_{\text{ac}}}{2\rho_{\text r} V\hbar}\frac{\text{d}}{\text{d}t} \left\vert \sum\limits_\qqq \text{i}q \sqrt{\frac{n_\qqq}{\omega_{\qqq}} }\int\limits^t_0 \text{d}t\, \re^{\text{i}\frac{ \varepsilon_\kkkp-\varepsilon_\kkk-\hbar \omega_{\qqq}}{\hbar}t} \int\limits_V \text{d}\rrr \psi^*_{\kkkp}(\rrr) \re^{ \text{i}\qqq\rrr} \psi_\kkk(\rrr)\right\vert^2,
\end{equation}
\item the probability of electron transition with phonon emission can be presented as follows:
\begin{equation}\label{eq-5}
W_e(\kkk, \kkkp)= \frac{D^2_{\text{ac}}}{2\rho_{\text r} V\hbar}\frac{\text{d}}{\text{d}t} \left\vert \sum\limits_\qqq \text{i}q \sqrt{\frac{n_\qqq +1}{\omega_{\qqq}} }\int\limits^t_0 \text{d}t\, \re^{\text{i}\frac{ \varepsilon_\kkkp-\varepsilon_\kkk+\hbar \omega_{\qqq}}{\hbar}t} \int\limits_V \text{d}\rrr \psi^*_{\kkkp}(\rrr) \re^{- \text{i}\qqq\rrr} \psi_\kkk(\rrr)\right\vert^2 .
\end{equation}
\end{itemize}
Here, $\psi_\kkk(\rrr, t)= \psi_\kkk(\rrr) \exp(-\text{i}\varepsilon_\kkk t/\hbar)$, $\psi_\kkk(\rrr)$
is the conduction electron wave function of stationary state $\kkk$, $\omega_{\qqq}$ is the phonon
angular frequency, $n_\qqq$ is the occupation number of the equilibrium phonons, which is given by
the Bose-Einstein distribution:
\begin{equation}\label{eq-6}
n_\qqq=1\left/\left[\exp (\hbar \omega_{\qqq} /k_{\text B}T)-1\right]\right..
\end{equation}
The summations in \eqref{eq-4} and \eqref{eq-5} should be carried out in the range of the first Brillouin zone (BZ).

After integration over $t$,  \eqref{eq-4} and \eqref{eq-5} are expressed as follows:
\begin{equation}\label{eq-7}
W_{a,e}(\kkk, \kkkp)= \frac{D^2_{\text{ac}}\hbar }{2\rho_{\text r} V }
\frac{\text{d}}{\text{d}t} \left\vert \sum\limits_\qqq
\frac{q}{\sqrt{\omega_{\qqq}}}\sqrt{n_\qqq+\frac{1}{2}\mp \frac{1}{2}}
\frac{\re^{\text{i}\frac{\varepsilon_\kkkp -\varepsilon_\kkk \mp \hbar \omega_{\qqq}}{\hbar} t}-1}{\varepsilon_\kkkp -\varepsilon_\kkk \mp \hbar \omega_{\qqq}}
\int\limits_V \text{d}\rrr \psi^*_{\kkkp}(\rrr) \re^{\pm \text{i}\qqq\rrr} \psi_\kkk(\rrr)\right\vert^2 .
\end{equation}
Here and bellow, upper and lower symbols of double signs refer to electron transitions with the $\qqq$ phonon absorption and emission, respectively.

In the presence of a uniform electric field ${\bf F}$ (which is parallel to the $z$-axis), the electron wave-function is determined by the stationary Schr\"{o}dinger equation which in the effective mass approximation is expressed as follows:
\begin{equation}\label{eq-8}
\left[-\frac{\hbar^2}{2m} \left(\frac{\text{d}^2}{\text{d}x^2}+\frac{\text{d}^2}{\text{d}y^2}+\frac{\text{d}^2}{\text{d}z^2}\right)+e Fz-\varepsilon_\kkk\right] \psi_\kkk (\rrr)=0.
\end{equation}
Here and below, the index $\kkk$ of $\varepsilon_\kkk$ is omitted to simplify the expressions, so we can substitute $\varepsilon_\kkk\to\varepsilon$ and $\varepsilon_{\kkkp}\to\varepsilon^\prime$  in further expressions.

The solution of \eqref{eq-8} is given by (see, e.g., \cite{1,7,8})
\begin{equation}\label{eq-9}
\psi_\kkk(\rrr)=\frac{1}{\sqrt{L_x L_y}} \, \re^{\text{i}(k_x x+k_y y)}\chi_n(z).
\end{equation}
Inserting \eqref{eq-9} into \eqref{eq-8}, the following equation is obtained \cite{1,7,8}
\begin{equation}\label{eq-10}
\left[-\frac{\hbar^2}{2m} \frac{\text{d}^2}{\text{d}z^2}+e Fz-\varepsilon_n\right]\chi_n(z)=0 .
\end{equation}
The solution of this equation is \cite{7} as follows:
\begin{equation}\label{eq-11}
\chi_n(z)=C_n \Ai \left[\frac{z}{l}-(k_n l)^2\right],
\end{equation}
where $C_n$ is the normalization constant, $l=(\hbar^2 /2eFm)^{1/3}$, $\Ai$ is the Airy function \cite{8}:

\begin{equation}\label{eq-12}
\Ai(s)=\frac{1}{\pi}\int\limits^\infty_0 \text{d}u\cos \left(\frac{u^3}{3}+us\right).
\end{equation}
Electron energy eigenvalues $\varepsilon$ are determined as follows:
\begin{equation}\label{eq-13}
\varepsilon=\frac{\hbar^2 k^2_{\perp}}{2m}+\varepsilon_n \,,
\end{equation}
where $\varepsilon_n=\hbar^2 k^2_n/2m$, the index $n$ identifies the electron energy eigenvalues, $k^2_\perp =k^2_x+k^2_y$, $\kkk_\perp$ is the electron wave-vector perpendicular to the electric field. Energy eigenvalues $\varepsilon_n$ (or $k_n$) are determined from the boundary conditions (see below) for the wave function $\chi_n(z)$.

For a semiconductor of length $L_z$ in $z$-direction (i.e., $-L_z/2\leqslant z\leqslant L_z/2$) from \eqref{eq-9} and relation \cite{1,3}
\begin{equation}\label{eq-14}
\frac{1}{L_z} \int\limits^{L_z/2}_{-L_z/2}\text{d}z\, \re^{\text{i}(q_z-q_{1, z})z}=\delta_{q_z, q_{1,z}}
\end{equation}
it follows
\begin{equation}\label{eq-15}
\int \text{d}\rrr^3 \psi^*_{\kkkp}(\rrr)\psi_{\kkk}(\rrr) \re^{\pm \text{i}\qqq\rrr}=
\delta_{k^\prime_x, k_x\pm q_x} \delta_{k^\prime_y, k_y\pm q_y}
\int\limits^{L_z/2}_{-L_z/2}\text{d}z\, \chi^*_{n^\prime}(z) \chi_n(z) \re^{\pm \text{i} q_z z}.
\end{equation}
Inserting \eqref{eq-15} into \eqref{eq-7} yields
\begin{multline}\label{eq-16}
W_{a,e}(\kkk, \kkkp)= \frac{D^2_{\text{ac}}\hbar }{2\rho_{\text r} V }\frac{\text{d}}{\text{d}t}
\left\vert \sum\limits_\qqq
\frac{q}{\sqrt{\omega_\qqq}}
\sqrt{n_\qqq+\frac{1}{2}\mp \frac{1}{2}}\right.\\
\times \left.
\frac{e^{\text{i}\frac{\varepsilon_\kkkp- \varepsilon_{\kkk}\mp \hbar \omega_\qqq}{\hbar} t} -1}{
\varepsilon_\kkkp- \varepsilon_\kkk\mp \hbar \omega_\qqq}
\delta_{k^\prime_x, k_x\pm q_x} \delta_{k^\prime_y, k_y\pm q_y}
\int\limits^{L_z/2}_{-L_z/2} \text{d}z\, \chi^*_{n^\prime}(z) \chi_n(z) e^{\pm \text{i} q_z z}\right\vert^2 .
\end{multline}
The Kronecker $\delta$ in this equation expresses the laws of conservation of perpendicular to the electric field quasi-momentum $x$, $y$ components of the scattered particles. Here,  only the normal N-processes of scattering are considered.

After summation with respect to $q_x$ and $q_y$ with the help of the Kronecker $\delta$, \eqref{eq-16} becomes as follows:
\begin{eqnarray}\label{eq-17}
W_{a,e}(\kkk, \kkkp)&=& \frac{D^2_{\text{ac}}\hbar }{2\rho_{\text r} V }\frac{\text{d}}{\text{d}t}
\left\vert \sum\limits_{q_z}
\left[ \frac{q}{\sqrt{\omega_\qqq}}
\sqrt{n_\qqq+\frac{1}{2}\mp \frac{1}{2}} %\right. \right. \\ \left. \left. \times
\frac{e^{\text{i}\frac{\varepsilon_\kkkp- \varepsilon_\kkk\mp \hbar \omega_\qqq}{\hbar} t}-1}{
\varepsilon_\kkkp-\varepsilon_\kkk\mp \hbar \omega_\qqq}\right]_{
\renewcommand{\arraystretch}{0.8}%
\scriptsize{\begin{array}{l} q_x=\pm (k^\prime_x -k_x),\\ q_y=\pm (k^\prime_y -k_y) \end{array}}}
\right.\nonumber\\
&&{}
\left.
{}\times
\int\limits^{L_z/2}_{-L_z/2}\text{d}z\, \chi^*_{n^\prime}(z) \chi_n(z) e^{\pm \text{i} q_z z}\right\vert^2 .
\end{eqnarray}
Taking into account that $\omega_\qqq$ is an even function on $\qqq$  (and particularly on $q_z$), from \eqref{eq-17} one obtains
\begin{eqnarray}\label{eq-18}
W_{a,e}(\kkk, \kkkp)&=& \frac{2D^2_{\text{ac}}\hbar }{\rho_{\text r} V }\frac{\text{d}}{\text{d}t}
\left\vert
\sum\limits_{q_z\geqslant 0}
\left[ \frac{q}{\sqrt{\omega_\qqq}}
\sqrt{n_\qqq+\frac{1}{2}\mp \frac{1}{2}}\;  %\right. \right. \\ \left. \left. \times
\frac{e^{\text{i}\frac{\varepsilon_\kkkp- \varepsilon_\kkk\mp \hbar \omega_\qqq}{\hbar} t}-1}{
\varepsilon_\kkkp-\varepsilon_\kkk\mp \hbar \omega_\qqq}\right]_{
\renewcommand{\arraystretch}{0.8}%
\scriptsize{\begin{array}{l} q_x=\pm (k^\prime_x -k_x),\nonumber\\ q_y=\pm (k^\prime_y -k_y) \end{array}}}
\right.
\\
&&{}\left. \times
\text{Re}
\int\limits^{L_z/2}_{-L_z/2}\text{d}z\, \chi^*_{n^\prime}(z) \chi_n(z) e^{\pm \text{i} q_z z}\right\vert^2 .
\end{eqnarray}
Using the following formal transformations
\begin{equation}\label{eq-19}
\frac{\text{d}}{\text{d}t} \left\vert \sum\limits_\qqq c_\qqq(t) \right\vert^2
=\frac{\text{d}}{\text{d}t} \sum\limits_{\qqq, \qqq_1} c_\qqq(t) c^*_{\qqq_1}(t)=
2\text{Re}\sum\limits_{\qqq, \qqq_1} \frac{\text{d}c_\qqq(t)}{\text{d}t} c^*_{\qqq_1}(t),
\end{equation}
\eqref{eq-18} may be written as follows:
\begin{eqnarray}\label{eq-20}
W_{a,e}(\kkk, \kkkp)&=& \frac{4D^2_{\text{ac}}}{\rho_{\text r} V }
\sum\limits_{
\renewcommand{\arraystretch}{0.7}%
\scriptsize{\begin{array}{c} q_z\geqslant 0\\ q_{1, z}\geqslant 0\end{array}}}
\text{Re}
\left[
\vphantom{\frac{
\re^{\text{i}(\mp \omega_\qqq\pm \omega_{\qqq_1})t}-
\re^{\text{i}\frac{\varepsilon_\kkkp- \varepsilon_\kkk\mp \hbar \omega_\qqq}{\hbar} t}
}{
\varepsilon_\kkkp-\varepsilon_\kkk\mp \hbar \omega_{\qqq_1}
}}
\frac{\text{i}qq_1}{\sqrt{\omega_\qqq \omega_{\qqq_1}}}
\sqrt{n_\qqq+\frac{1}{2}\mp \frac{1}{2}}
\sqrt{n_{\qqq_1}+\frac{1}{2}\mp \frac{1}{2}} \; \right. \nonumber\\ &&{}\left. \times
\frac{
\re^{\text{i}(\mp \omega_\qqq\pm \omega_{\qqq_1})t}-
\re^{\text{i}\frac{\varepsilon_\kkkp- \varepsilon_\kkk\mp \hbar \omega_\qqq}{\hbar} t}
}{
\varepsilon_\kkkp-\varepsilon_\kkk\mp \hbar \omega_{\qqq_1}
}\right]_{
\renewcommand{\arraystretch}{0.8}%
\scriptsize{
\begin{array}{l} q_x=\pm (k^\prime_x -k_x),\\ q_y=\pm (k^\prime_y -k_y) \end{array}}
}\nonumber\\&&{}\times
\text{Re}
\int\limits^{L_z/2}_{-L_z/2}\text{d}z\,
\chi^*_{n^\prime}(z) \chi_n(z) \re^{\pm \text{i} q_z z}
\text{Re}
\int\limits^{L_z/2}_{-L_z/2}\text{d}z_1\,
\chi^*_{n^\prime}(z_1) \chi_n(z_1) \re^{\pm \text{i} q_{1, z} z_1}.
\end{eqnarray}
For  a sufficiently long time $t$, when the relation $\lim\limits_{t\to\infty} \sin(at)/a =\pi \delta(a)$ can be used \cite{1,2},  \eqref{eq-20} is expressed as follows:
\begin{eqnarray}\label{eq-21}
W_{a,e}(\kkk, \kkkp)&=& \frac{4D^2_{\text{ac}}}{\rho_{\text r} V }
\sum\limits_{
\renewcommand{\arraystretch}{0.7}%
\scriptsize{\begin{array}{c} q_z\geqslant 0\\ q_{1, z}\geqslant 0 \end{array}}}
\left\{
\frac{qq_1}{\sqrt{\omega_{\qqq} \omega_{\qqq_1}}}
\sqrt{n_\qqq+\frac{1}{2}\mp \frac{1}{2}}
\sqrt{n_{\qqq_1}+\frac{1}{2}\mp \frac{1}{2}}\;
 \right.\nonumber\\
&&{}\left. \times\left[-\pi
\frac{
(\pm \omega_{\qqq}\mp \omega_{{\qqq}_1})\delta (\omega_{\qqq} -\omega_{\qqq_1})}{
\varepsilon_\kkkp- \varepsilon_\kkk \mp \hbar \omega_{{\qqq}_1}}
+\pi \delta (\varepsilon_\kkkp- \varepsilon_\kkk\mp \hbar \omega_{\qqq})
\frac{
\varepsilon_\kkkp- \varepsilon_\kkk\mp \hbar \omega_{\qqq}
}{
\varepsilon_\kkkp- \varepsilon_\kkk\mp \hbar \omega_{{\qqq}_1}
}
\right]
\vphantom{\sqrt{n_\qqq+\frac{1}{2}\mp \frac{1}{2}}}
\right\}_{
\renewcommand{\arraystretch}{0.8}%
\scriptsize{\begin{array}{l} q_x=\pm (k^\prime_x -k_x),\\ q_y=\pm (k^\prime_y -k_y) \end{array}}}
\nonumber\\
&&{}
\times
\text{Re}
\int\limits^{L_z/2}_{-L_z/2}\text{d}z\, \chi^*_{n^\prime}(z) \chi_n(z) \re^{\pm \text{i} q_z z}
\text{Re} \int\limits^{L_z/2}_{-L_z/2}\text{d}z_1\, \chi^*_{n^\prime}(z_1) \chi_n(z_1) \re^{\pm \text{i} q_{1, z} z_1} .
\end{eqnarray}
In this expression, the summation with respect to $q_{1, z}$ is non-zero only when  $q_z=q_{1, z}$. Then, electron transition probability can be written as follows:
\begin{eqnarray}\label{eq-22}
W_{a,e}(\kkk, \kkkp)&=& \frac{4\pi D^2_{\text{ac}}}{\rho_{\text r} V }
\sum\limits_{
\renewcommand{\arraystretch}{0.7}%
\scriptsize{\begin{array}{c} q_x, q_y, \\ q_z\geqslant 0\end{array}}}
\frac{q^2}{\omega_{\qqq} }
\left(n_{\qqq}+\frac{1}{2}\mp \frac{1}{2}\right) \delta
\left(\varepsilon_\kkkp- \varepsilon_\kkk\mp \hbar \omega_{\qqq}\right)
\nonumber\\
&&{}
\times \delta_{k^\prime_x, k_x\pm q_x}\delta_{k^\prime_y, k_y\pm q_y}
\left[\text{Re}
\int\limits^{L_z/2}_{-L_z/2}\text{d}z\, \chi^*_{n^\prime}(z) \chi_n(z) \re^{\pm \text{i} q_z z}\right]^2 .
\end{eqnarray}
Here, the Dirac $\delta$-function indicates the energy conservation law.

Note,  if in  \eqref{eq-22}  the plane wave ($\chi_n(z)= \re^{\text{i} k_zz}/\sqrt{L_z}$) or Bloch function is taken as an electron wave function,  then the well-known classical expression of probability  $W_{a,e}(\kkk, \kkkp)$ \cite{1,2,3,4} (Fermi Golden rule) is derived. In the common case, the calculation of the integral and sum in  \eqref{eq-22} is complicated. On the other hand, the expression  \eqref{eq-1} for $\tau_{\kkk, \text{ac}}$ is derived (within the framework of the flat-band semiconductor model) by using the approximation of elasticity of electron-acoustic phonon scattering (elastic collision approximation, $\hbar \omega_\qqq\ll \varepsilon_\kkk$) with the assumption that
\begin{equation}\label{eq-23}
n_\qqq+1\approx n_\qqq\cong \frac{k_{\text B} T}{\hbar\omega_{\qqq}} =
\frac{k_{\text B} T}{\hbar \nu_0 q}\gg 1\, .
\end{equation}
Here, $\omega_\qqq=\nu_0 q$ dispersion law of the long wavelength longitudinal acoustic phonon is used.

To reveal the difference between the results of flat and titled band approaches it is reasonable that these assumptions should be used here as well. Then, after simple summing with respect to $q_x$  and $q_y$,  \eqref{eq-22} takes the following simpler form:
\begin{equation}\label{eq-24}
W_{a,e}(\kkk, \kkkp)=
\frac{4\pi D^2_{\text{ac}}k_{\text B} T}{\rho_{\text r} V\hbar \nu^2_0 }\delta (\varepsilon_\kkkp -\varepsilon_\kkk)
\sum\limits_{q_z\geqslant 0}
\left[\text{Re}
\int\limits^{L_z/2}_{-L_z/2}\text{d}z\, \chi^*_{n^\prime}(z) \chi_n(z) \re^{\pm \text{i} q_z z}\right]^2 .
\end{equation}
In what follows, we use the transformation
\begin{eqnarray}\label{eq-25}
\lefteqn{\sum\limits_{q_z\geqslant 0}
\left[\text{Re}\int\limits^{L_z/2}_{-L_z/2}\text{d}z\, \chi^*_{n^\prime}(z) \chi_n(z) \re^{\pm \text{i} q_z z}\right]^2 }\qquad \nonumber\\
&&{}= \sum\limits_{q_z\geqslant 0}
\left[\frac{1}{2}\int\limits^{L_z/2}_{-L_z/2}\text{d}z
\left[\chi^*_{n^\prime}(z) \chi_n(z) \re^{\pm \text{i} q_z z}+\chi_{n^\prime}(z) \chi^*_n(z) \re^{\mp \text{i} q_z z}\right] \right]^2 \nonumber\\
&&{}=\frac{1}{2}\text{Re} \sum\limits_{q_z\geqslant 0} \int\limits^{L_z/2}_{-L_z/2}
\int\limits^{L_z/2}_{-L_z/2} \text{d}z\, \text{d}z_1
\left[\chi^*_{n^\prime}(z) \chi_n(z) \chi_{n^\prime}(z_1) \chi^*_n(z_1) \re^{\pm \text{i} q_z z\mp \text{i} q_z z_1}
\right.\nonumber\\[2ex]
&&\left. \qquad\qquad\qquad +\chi^*_{n^\prime}(z) \chi_n(z) \chi^*_{n^\prime}(z_1) \chi_n(z_1) \re^{\pm \text{i} q_z z\pm \text{i} q_z z_1}
\right]
\end{eqnarray}
and the relation \cite{3}
\begin{equation}\label{eq-26}
\sum\limits_{q_z} \re^{\text{i}q_z (z-z_1)}=L_z \delta (z-z_1).
\end{equation}
The result is as follows:
\begin{eqnarray}\label{eq-27}
W_{a,e}(\kkk, \kkkp)&=&
\frac{\pi D^2_{\text{ac}}k_{\text B} T L_z}{\rho_{\text r} V\hbar \nu^2_0 }\delta
(\varepsilon_{\kkkp}-\varepsilon_\kkk) %\\\times
\text{Re}\int\limits^{L_z/2}_{-L_z/2}
\int\limits^{L_z/2}_{-L_z/2} \text{d}z\, \text{d}z_1
\left[\chi^*_{n^\prime}(z) \chi_n(z) \chi_{n^\prime}(z_1) \chi^*_n(z_1) \delta(z-z_1)
\right.\nonumber\\[2ex]
&&{}
\left. +\chi^*_{n^\prime}(z) \chi_n(z) \chi^*_{n^\prime}(z_1) \chi_n(z_1) \delta(z+z_1)
\right].
\end{eqnarray}
Delta-integration over $z_1$ yields:
\begin{eqnarray}\label{eq-28}
W_{a,e}(\kkk, \kkkp)&=&
\frac{\pi D^2_{\text{ac}}k_{\text B} T L_z}{\rho_{\text r} V\hbar \nu^2_0 }\delta
(\varepsilon_{\kkkp}-\varepsilon_\kkk) \nonumber\\
&&{}
\times \text{Re}\int\limits^{L_z/2}_{-L_z/2}
\text{d}z\left\{
\left\vert\chi_{n^\prime}(z) \right\vert^2\left\vert\chi_{n}(z) \right\vert^2+
\chi^*_{n^\prime}(z) \chi_{n}(z) \chi^*_{n^\prime}(-z) \chi_n(-z) \right\}.
\end{eqnarray}

\begin{figure}[htbp]
\centerline{\includegraphics[height=60mm]{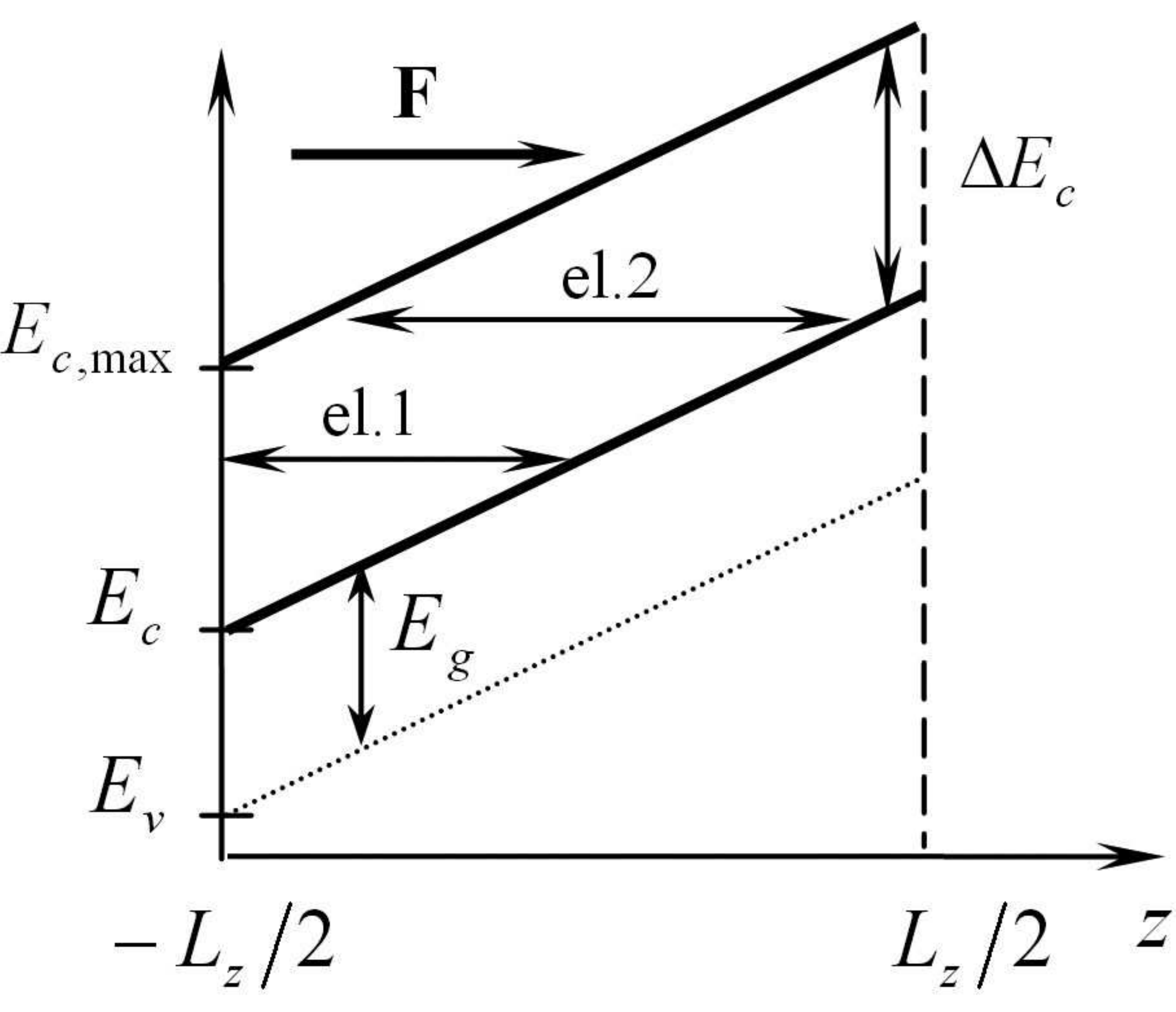}}
\caption{Semiconductor energy-band diagram in
the presence of a uniform electric field   $F$
(parallel to the $z$-axis); el.2 -- Bloch oscillations.
}
\label{fig1}
\end{figure}

Electron energy eigenvalues $\varepsilon_n$ are determined from boundary conditions to  \eqref{eq-10}. Note, there is some difference between the peculiarities of the movement of an electron under electric field in vacuum and an electron in a semiconductor. Conduction band gap $\Delta E_{\text{c}}$ of a semiconductor is a finite quantity. Contrary to the vacuum, the movement of a conduction electron in a semiconductor with perfect lattice has an oscillation character, as it is shown in figure~\ref{fig1}.

Electron `el.1' oscillates between the bottom of the conduction band and semiconductor edge; electron `el.2' periodically reflects from the conduction band top and bottom edges. The behavior of an `el.2' is well-known as a Bloch oscillation. Boundary conditions for Schr\"{o}dinger equation \eqref{eq-10} and, therefore, electron energy eigenvalues in case of Bloch oscillations have been reported, for example, in \cite{9}.
Here, for definiteness, neglecting the Bloch oscillations, the case of `el.1' is considered only, i.e., it is assumed that the electric field is low and the magnitude of $\Delta E_{\text{c}}$ is very large. In other words, the model of a triangular quantum well with finite sizes is considered.

Here, we are interested in a large length semiconductor, particularly, in $z$-direction. Then, the allowed values of $k_n$ (or $\varepsilon_n$) are computed from the boundary condition $\chi_n$ $(z=-L_z/2)=~0$ as (see, \cite{7} and figure~\ref{fig1})
\begin{equation}\label{eq-29}
\frac{L_z}{2l}+(k_n l)^2=-a_n\,,
\end{equation}
where $a_n$ are the zeros of the Airy function which are located in the negative part of the real axis \cite{8}.
They are well approximated as $a_n\cong -(3\pi[4n-1]/8)^{2/3}$, where $n=1,2,\ldots\, $. Inserting this relation into
\eqref{eq-29} we obtain
\begin{equation}\label{eq-30}
\frac{L_z}{2l}+(k_n l)^2=\left[\frac{3\pi}{2}\left(n-\frac{1}{4}\right)\right]^{2/3}.
\end{equation}
Solving  \eqref{eq-30} for $k_n$ and inserting the result into  \eqref{eq-13}, one obtains energy eigenvalues \cite{7}:
\begin{equation}\label{eq-31}
\varepsilon=\frac{\hbar^2}{2m}\left\{k^2_\perp -\frac{L_z}{2l^3}+\frac{1}{l^2}\left[ \frac{3\pi}{2} (n-1/4)\right]^{2/3}\right\}.
\end{equation}
The values $\varepsilon$ in  \eqref{eq-31} are obtained as a function of the electron state quantum numbers.
The quantities $k_x$, $k_y$ (or $\kkk_\perp$) and $n$ (or $k_n$) are a set of quantum numbers which determine
the conduction electron state in the presence of an electric field.

Normalization constant $C_n$ in  \eqref{eq-11} is determined as follows:
\begin{equation}\label{eq-32}
C^{-2}_n=\int\limits^{L_z/2}_{-L_z/2} \text{d}z\, \left\vert \Ai\left(\frac{z}{l} -(k_nl)^2\right) \right\vert^2.
\end{equation}
It is known \cite{7,8} that
\begin{equation}\label{eq-33}
\int \text{d}s\, \Ai^2(s)=s\Ai^2(s)-\Ai^{\prime\, 2}(s),
\end{equation}
where $\Ai^{\prime}$ is the derivative of the Airy function.

Consequently, normalization constant $C_n$ can be presented as follows:
\begin{equation}\label{eq-34}
C^{-2}_n=l\left. \left[s \Ai^2(s)-\Ai^{\prime\, 2}(s)\right]\right\vert^{L_z/l+a_n}_{a_n}.
\end{equation}
For a semiconductor with large length $L_z$ one has
\begin{equation}\label{eq-35}
C^{-2}_n=l \Ai^{\prime\, 2}(a_n).
\end{equation}
Here, we used the fact that $\Ai(a_n)=0$; functions $\Ai(s)$ and $\Ai^\prime(s)$ exponentially vanish for
positive large argument $s$ \cite{8}. On the other hand, the value $\Ai^{\prime}(a_n)$ is well approximated as \cite{8}
\begin{equation}\label{eq-36}
\Ai^\prime(a_n)\cong(-1)^{n-1}\frac{1}{\sqrt{\pi}}(-a_n)^{1/4}.
\end{equation}
Insertion of  \eqref{eq-36} into  \eqref{eq-35} yields:
\begin{equation}\label{eq-37}
C^2_n =\pi/l{\sqrt{-a_n}} \,.
\end{equation}
Thus, using \eqref{eq-11}, \eqref{eq-29} and taking into account that $\chi_n(z)$ is the real function
[see, \eqref{eq-11}, \eqref{eq-12}] the expression \eqref{eq-28} can be presented as follows:
\begin{eqnarray}\label{eq-38}
W_{a,e}(\kkk, \kkkp)&=&
\frac{\pi D^2_{\text{ac}}k_{\text B} T L_zl}{\rho_{\text r} V\hbar \nu^2_0 }\delta
(\varepsilon_{\kkkp}-\varepsilon_\kkk) C^2_{n^\prime} C^2_n
\left\{ \int\limits^{L_z/2l}_{-L_z/2l} \text{d}u\, \left\vert
\Ai\left(u-(k_{n^{\prime} }l)^2\right)
\right\vert^2
\left\vert
\Ai\left(u-(k_{n }l)^2\right)
\right\vert^2 \right. \nonumber\\
&&{}
\left.  + \int\limits^{L_z/2l}_{-L_z/2l} \text{d}u\, \left[
\Ai\left(u-(k_{n^\prime }l)^2\right) \Ai\left(-u-(k_{n^\prime }l)^2\right)
\Ai\left(u-(k_{n}l)^2\right) \Ai\left(-u-(k_{n }l)^2\right)
\right] \right\},
\end{eqnarray}
where $u=z/l$ is the dimensionless variable of integration.
 \eqref{eq-38} describes electron transition probability in the presence of an electric field.
 Transition probability depends on the electric field and it is a symmetric function: $W_{a,e}(\kkk, \kkkp)= W_{a,e}(\kkkp, \kkk)$.

%%%%%%%%%%%%%%%%%%%%%%%%%%%%%%%%

\section{Electron-acoustic phonon FIT scattering time}

 The scattering time is defined by \cite{3}:
\begin{equation}\label{eq-39}
\tau^{-1}_{\kkk, \text{sc}}=\sum\limits_{\kkkp} W(\kkk, \kkkp)
\end{equation}
 \eqref{eq-38} shows that in the present case, the probabilities of phonon absorption and emission
 by electron are the same: $W_{a}(\kkkp, \kkk)= W_{e}(\kkk, \kkkp)$. Then,  \eqref{eq-39} can be written as follows:
\begin{equation}\label{eq-40}
\tau^{-1}_{\kkk, \text{sc}}=2\sum\limits_{\kkkp} W_e(\kkk, \kkkp).
\end{equation}
In the common case, the scattering time (the inverse of the scattering rate) differs from the relaxation time although
sometimes both of them are equivalent (for example, the above-mentioned case of  \eqref{eq-1}) \cite{3}.
To calculate  $\tau^{-1}_{\kkk, \text{sc}}$ in  \eqref{eq-40}, we replace the sums
over $k^{\prime}_x$  and $k^{\prime}_y$ by integrals over $k^{\prime}_x$  and $k^{\prime}_y$, respectively.
The transition from the sum to integral with the help of the relation $\text{d}k^{\prime}_x \, \text{d}k^{\prime}_y=\pi\,
\text{d}k^{\prime 2}_\perp$ can be presented as follows:
\begin{equation}\label{eq-41}
\sum\limits_{\kkkp}\longrightarrow\frac{2L_x L_y}{(2\pi)^2} \int\limits_{\text{BZ}} \text{d}k^{\prime}_x \, \text{d}k^{\prime}_y
\sum\limits_{n^\prime}=\frac{L_x L_y}{2\pi} \int\limits_{\text{BZ}} \text{d}k^{\prime\, 2}_\perp \sum\limits_{n^\prime}\, . \end{equation}
Here, coefficient 2 in the numerator is related to the electron spin.

%%%%%%%%%%%%%%%%%%%%%%%%%%%%%%%%
From  \eqref{eq-30} it follows that $k^2_{n+1}-k^2_n \sim 1/l^2$. The distance between $k_n$ and $k_{n+1}$
depends on $n$ and it is small for large $l$. Therefore, at low-field regime, one can change the summation
over $n^\prime$ by an integral over $k_{n^\prime}$:
\begin{equation}\label{eq-42}
\sum\limits_{n^\prime}\longrightarrow \int \text{d}k_{ n^\prime}
\frac{\text{d} n^\prime}{\text{d}k_{ n^\prime}}\, .
\end{equation}
The derivative $\text{d} n^\prime/ \text{d}k_{n^\prime}$ can be evaluated based on \eqref{eq-30}.
The solution of  \eqref{eq-30} for $n$ is given the following expression for $n^\prime$:
\begin{equation}\label{eq-43}
n^\prime=\frac{2}{3\pi} \left(\frac{L_z}{2l}+(k_{ n^\prime} l)^2\right)^{3/2}+\frac{1}{4}\, .
\end{equation}
Therefore,
\begin{equation}\label{eq-44}
\frac{\text{d} n^\prime}{\text{d}k_{n^\prime}}=\frac{2l^2 k_{ n^\prime} }{\pi} \sqrt{\frac{L_z }{2l}+(k_{ n^\prime} l)^2}\, .
\end{equation}
Based on \eqref{eq-29}, the derivative $\text{d} n^\prime/ \text{d}k_{n^\prime}$ can be presented as follows:
\begin{equation}\label{eq-45}
\frac{\text{d} n^\prime}{\text{d}k_{n^\prime}}=\frac{2l^2 k_{ n^\prime} }{\pi} \sqrt {-a_{n^\prime}}\, .
\end{equation}
Thus, transition  \eqref{eq-41} can be presented as follows:
\begin{equation}\label{eq-46}
\sum\limits_{n^\prime}\longrightarrow \frac{L_x L_y l^2}{2\pi^2}
\int\limits^{k^{\prime 2}_{\perp, \max}}_{k^{\prime 2}_{\perp, \min}}
\int\limits^{k^2_{n^\prime, \max}}_{k^2_{n^\prime, \min}}
\text{d} k^{\prime\, 2}_\perp \text{d} k^2_{n^\prime}\sqrt {-a_{n^\prime}}\, .
\end{equation}
Inserting  \eqref{eq-38} into  \eqref{eq-40}, simultaneously taking into account  \eqref{eq-37}
and transition \eqref{eq-46}, for the scattering time one obtains:
\begin{eqnarray}\label{eq-47}
\tau^{-1}_{\kkk , \text{sc}}&=&
\frac{D^2_{\text{ac}} k_{\text B} TC^2_n}{\rho_{\text r} \hbar \nu^2_0}
\int\limits^{k^{\prime 2}_{\perp, \max}}_{k^{\prime 2}_{\perp, \min}}
\int\limits^{s^{\prime}_{\max}}_{s^{\prime}_{\min}}
\text{d} k^{\prime\, 2}_\perp \text{d}s^\prime %\\\times
\delta (\varepsilon_\kkkp -\varepsilon_\kkk)
\left\{
\int\limits^{L_z/2l}_{-L_z/2l} \text{d}u\, \left\vert \Ai(u-s^\prime)\right\vert^2
\left\vert \Ai(u-(k_n l)^2)\right\vert^2
\right. \nonumber\\
&&{}
\left. +\int\limits^{L_z/2l}_{-L_z/2l}
\text{d}u\,
\Ai(u-s^\prime)\, \Ai(-u-s^\prime)\,
\Ai(-u-(k_n l)^2)\, \Ai(u-(k_n l)^2)
\right\}\, .
\end{eqnarray}
Here, $s^\prime=(lk_{n^\prime})^2$ is the dimensionless variable of integration, Brillouin zone
is replaced by the infinite range: $0\leqslant k^{\prime 2}_\perp <\infty$. The limits of
integration over $s^\prime$ are determined by the $\delta$-function
\begin{equation}\label{eq-48}
\delta (\varepsilon_{\kkkp} -\varepsilon_{\kkk})=
\frac{2m}{\hbar^2} \, \delta \left(k^{\prime \, 2}_\perp + k^2_{n^\prime}-k^2_\perp -k^2_n \right),
\end{equation}
as:  $s^{\prime}_{\min} =-a_1 -L_z/2l$, $s^{\prime}_{\max}= \left(k^2_\perp -k^2_n \right) l^2$,
where $a_1=-(9\pi/8)^{2/3}$ is the first zero of the Airy function \cite{8}.

$\delta$-integration over $s^\prime$ in  \eqref{eq-47} yields
\begin{eqnarray}\label{eq-49}
\tau^{-1}_{\kkk , \text{sc}}&=&\frac{2D^2_{\text{ac}}m k_{\text B} TC^2_n}{\rho_{\text r} \hbar^3 \nu^2_0}
\int\limits^\infty_0 \text{d}s^\prime_\perp\left\{
\int\limits^{L_z/2l}_{-L_z/2l} \text{d}u\, \left\vert \Ai(u+s^\prime_\perp -\Delta)\right\vert^2
\left\vert \Ai(u-(k_n l)^2)\right\vert^2 \right. \nonumber\\
&&{}
\left. +\int\limits^{L_z/2l}_{-L_z/2l} \text{d}u\,
\Ai(u+s^\prime_\perp-\Delta)\, \Ai(-u+s^\prime_\perp-\Delta)\,
\Ai(-u-(k_n l)^2)\, \Ai(u-(k_n l)^2)\right\}\, .
\end{eqnarray}
Here, $s^\prime_\perp =k^{\prime\, 2}_\perp l^2$ is the dimensionless variable of integration,
$\Delta\equiv k^2_\perp l^2+k^2_n l^2=2m\varepsilon l^2 / \hbar^2$ is the dimensionless energy of an electron.

In  \eqref{eq-49}, evaluations of the first and second the integrals over $s^\prime_\perp$ can be carried out
based on \eqref{eq-33} and following the integral of the product of two Airy functions \cite{8}, respectively,
\begin{equation}\label{eq-50}
\int\limits^\infty_0 \text{d}s\, \Ai(b+s)\, \Ai(c+s)=\frac{1}{b-c} \left[\Ai(b) \Ai^\prime (c)-\Ai^\prime(b) \Ai(c) \right].
\end{equation}
The result is as follows:
\begin{equation}\label{eq-51}
\tau^{-1}_{\kkk , \text{sc}}=\frac{D^2_{\text{ac}}m k_{\text B} TC^2_n}{\rho_{\text r} \hbar^3 \nu^2_0}[I_1+I_2],
\end{equation}
where
 \begin{equation}\label{eq-52}
 I_1=\int\limits^{L_z/2l}_{-L_z/2l} \text{d}u\,
 \left\{(\Delta -u) \Ai^2(u-\Delta)+\Ai^{\prime\, 2}(u-\Delta)\right\}
\left\vert \Ai(u-(k_n l)^2)\right\vert^2,
\end{equation}
\begin{multline}\label{eq-53}
I_2=\int\limits^{L_z/2l}_{-L_z/2l} \text{d}u\, \frac{1}{2u} \left[
\Ai(u-\Delta) \Ai^\prime(-u-\Delta)%
 - \Ai^\prime (u-\Delta)\Ai(-u-\Delta)\right]
\Ai\left(-u-(k_n l)^2\right)\, \Ai\left(u-(k_n l)^2\right) .
\end{multline}
The functional analyses of the sub-integral expression show that the main contribution in integrals \eqref{eq-52}
and \eqref{eq-53} is given by the range near $u\sim 0$. An approximate estimation of the integral \eqref{eq-53} is
carried out with the help of L'Hopital's rule. The result is as follows:
\begin{equation}\label{eq-54}
I_1+I_2\approx \frac{1}{lC^2_n} \left[\Delta \Ai^2(-\Delta)+\Ai^{\prime \, 2}(-\Delta)\right].
\end{equation}
Therefore, the electron scattering time can be presented as follows:
\begin{equation}\label{eq-55}
\tau^{-1}_{\kkk , \text{sc}}=\frac{D^2_{\text{ac}}m k_{\text B} T}{\rho_{\text r} \hbar^3 \nu^2_0 l}
\left[\Delta \Ai^2(-\Delta)+\Ai^{\prime\, 2}(-\Delta)\right].
\end{equation}

\section{Summary}

The electron-acoustic phonon scattering theory has been reconsidered. In semiconductors,
whose bands are tilted under uniform electric field, the time of electron scattering by acoustic phonon
is determined by  \eqref{eq-55}. Scattering time depends on the electron energy $\varepsilon$. It depends on
the electric field as well, because $\Delta\sim \varepsilon l^2$; $l\sim F^{-1/3}$. Those dependencies are
determined by the Airy function properties \cite{8}. Thus, for negative arguments, the Airy function oscillates.
From the asymptotic series of the Airy function $\Ai(\Delta)$ and of their derivative $\Ai^\prime(\Delta)$ it
follows that for large negative argument \cite{8}
\begin{equation}\label{eq-56}
\vert\Delta\vert \Ai^2 (-\vert\Delta\vert)+\Ai^{\prime 2}(-\vert\Delta\vert)=\vert\Delta\vert^{1/2}/\pi\, .
\end{equation}
The Airy function decays exponentially for positive arguments. The first terms of asymptotic series of the Airy
function and of their derivative for positive arguments are as follows:
\begin{equation}\label{eq-57}
\Ai(\vert\Delta\vert) =\frac{1}{2\sqrt{\pi}\, \vert\Delta\vert^{1/4}} \exp\left(-2 \vert\Delta\vert^{3/2}/3\right), \qquad
\Ai^\prime(\vert\Delta\vert) =-\frac{\vert\Delta\vert^{1/4}}{2\sqrt{\pi}} \exp\left(-2 \vert\Delta\vert^{3/2}/3\right).
\end{equation}
Then, from \eqref{eq-55}, \eqref{eq-56} and \eqref{eq-57} it follows:
\begin{itemize}
\item for positive large $\Delta$
\begin{equation}\label{eq-58}
\tau^{-1}_{\kkk , \text{sc}}=\frac{D^2_{\text{ac}}m k_{\text B} T\vert\Delta\vert^{1/2}}{
\pi \rho_{\text r} \hbar^3 \nu^2_0 l}\, ,
\end{equation}
\item for negative large $\Delta$
\begin{equation}\label{eq-59}
\tau^{-1}_{\kkk , \text{sc}}=\frac{D^2_{\text{ac}}m k_{\text B} T\vert\Delta\vert^{1/2}}{
\pi \rho_{\text r} \hbar^3 \nu^2_0 l}\frac{1}{2}\exp\left(-4 \vert\Delta\vert^{3/2}/3\right).
\end{equation}
\end{itemize}
Insertion $\Delta$ and $l$ into  \eqref{eq-59} yields:
\begin{equation}\label{eq-60}
\tau^{-1}_{\kkk , \text{sc}}=\tau^{-1}_{\kkk , \text{ac}}
\frac{1}{2}\, \exp\left(-4 \sqrt{2m\vert\varepsilon\vert^3 / \hbar^2}\, /3 \re F\right).
\end{equation}
General expression \eqref{eq-55} for the electron-acoustic phonon scattering time can be modified
by inserting $\Delta$ into  \eqref{eq-55}. Then, one has
\begin{equation}\label{eq-61}
\tau^{-1}_{\kkk , \text{sc}}=\frac{D^2_{\text{ac}}m k_{\text B} T}{\rho_{\text r} \hbar^3 \nu^2_0 l}
\left[\frac{2m\varepsilon l^2}{\hbar^2}\, \Ai\left(-\frac{2m\varepsilon l^2}{\hbar^2}\right) +\Ai^{\prime\, 2}\left(-\frac{2m\varepsilon l^2}{\hbar^2}\right)\right].
\end{equation}

\begin{figure}[htbp]
\centerline{\includegraphics[height=60mm]{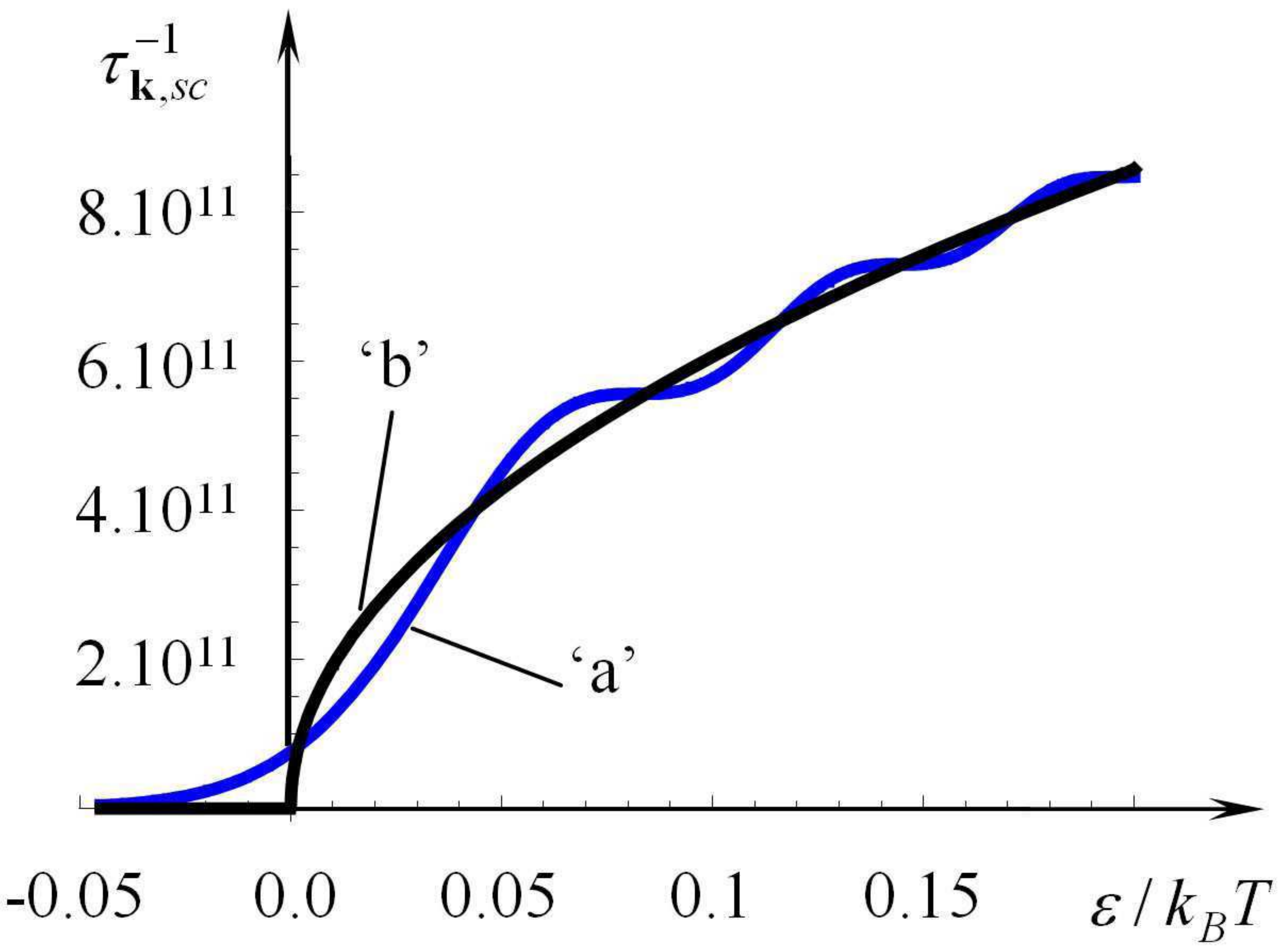}}
\caption{(Color online) The dependencies of $\tau^{-1}_{\kkk , \text{sc}}$  (curve `a' -- $F=800$~V/cm) and
$\tau^{-1}_{\kkk , \text{ac}}$  (curve `b' --  $F=0$) on  electron dimensionless energy $\varepsilon/k_{\text B}T$  for n-Si at $T=300$~K.}
\label{fig2}
\end{figure}

It is easy to establish that \eqref{eq-58} and \eqref{eq-1} are the same. As equation~\eqref{eq-60} shows,
the scattering time of low energy electron depends on the electric field. The dependence has exponential character.
This effect has been explained in terms of the penetration of the electron wave function into a
band gap of a semiconductor \cite{6}. As a result, for low energy electrons, which are located near the bottom edge of
the conduction band, the transition with phonon emission becomes allowable in the region below the conduction
band edge. Note, at flat-bands approach, there is a threshold of a phonon emission by a low energy electron,
for detailes see~\cite{10}. All these peculiarities are well displayed in figure~\ref{fig2}, where the dependencies
of $\tau^{-1}_{\kkk , \text{ac}}$ and $\tau^{-1}_{\kkk , \text{sc}}$ on electron dimensionless energy
$\varepsilon/k_{\text B}T$ are plotted for n-Si at $T=300$~K with the following parameters \cite{5}:
$m=0.32 m_0$, $\rho_{\text r}=2329$~kg/m$^3$, $\nu_0=8.43\cdot 10^3$~m/s, $D_{\text{ac}}=9$~eV.
The dependence $\tau^{-1}_{\kkk , \text{sc}}(\varepsilon/k_{\text B}T)$ (curve `a')
is calculated at $F=800$~V/cm based on \eqref{eq-61}. The dependence $\tau^{-1}_{\kkk , {\text ac}}(\varepsilon/k_{\text B}T)$
(curve `b') is calculated based on \eqref{eq-1}. As shown in figure~\ref{fig2}, the curve `a' has a character of
light oscillations around the curve `b'. At a low field regime $F<\sim 400$~V/cm, the curves `a' and `b' practically
coincide in the range of positive energy. Other important peculiarities are as follows: on the curve `b' there is a fracture,
i.e., $\left. \text{d}\tau^{-1}_{\kkk , \text{ac}}/\text{d}\varepsilon\right\vert_{\varepsilon =0}=\infty$
[see, equation~\eqref{eq-1}]; on the curve `a' there is no fracture [see,  \eqref{eq-61}].

Taking into acount the problem solution reported in \cite{6} it can be stated that the results of the present study
can have a principial effect on the mobility fluctuation theory,  especially. It should be
noted that in the above-presented electron-acoustic phonon FIT scattering study, the electron
quasi-momentum relaxation time and its relation to the scattering time is not included. It requires a separate consideration.

%%%%%%%%%%%%%%%%%%%%%%%%%%%%%%%%%%%

\newpage
\ukrainianpart

\title{Індуковане електронноакустичним фононним полем тунельне розсiяння}
\author{С.В. Мелконян, А.Л. Харатюнян, Т.А. Залінян}

\address{Факультет фізики напівпровідників і мікроелектроніки, Єреванський державний університет,
0025 Єреван, Вірменія}

\makeukrtitle
\begin{abstract}
У статтi подано новий погляд на теорiю електронноакустичного розсiювання
одного фонона. При цьому припускається, що невироджений напiвпровiдник має
сферичну параболiчну зонну
структуру. В основу перегляду теорiї покладено ефект нахилу напiвпровiдникових
зон при накладанні збурюючого потенцiалу електричного поля. У цьому випадку
власнi функцiї електрона вже не є плоскими хвилями чи функцiями Блоха. В
режимi слабких полів отримано вирази для ймовiрностi електронних
внутрішньозонних переходів i для часу розсiяння в наближеннi
пружніх зiткнень. Також проаналiзовано залежнiть часу розсiяння вiд енергiї
електрона та напруженості однорiдного електричного поля. Представлено
результати вiдповiдних числових обчислень для n-Si при температурi 300~K.
Встановлено відсутність зламу на кривій залежностi часу розсiяння електрона
вiд енергiї електрона.

\keywords напівпровідник з нахиленою зоною, електронноакустичне фононне розсіяння, ймовірність переходу, час розсіяння
\end{abstract}
\end{document}